\documentclass[a4paper,11pt]{article}

\usepackage{amsmath,amssymb}    
\usepackage{color}
\usepackage{cite}               
\usepackage{hyperref}           
\usepackage{multirow,makecell}   

\numberwithin{equation}{section}   

\def \be {\begin{equation}}
\def \ee {\end{equation}}
\def \ba {\begin{array}}
\def \ea {\end{array}}
\def \bea{\begin{eqnarray}}
\def \eea{\end{eqnarray}}

\def \a {\alpha}

\def \g {\gamma}

\def \G {\Gamma}
\def \d {\delta}
\def \D {\Delta}
\def \e {\epsilon}

\def \m {\mu}
\def \n {\nu}

\def \s {\sigma}

\def \r {\rho}

\def \mA {\mathcal A}
\def \mB {\mathcal B}

\def \mD {\mathcal D}

\def \mJ {\mathcal J}
\def \mK {\mathcal K}

\def \mO {\mathcal O}
\def \mP {\mathcal P}
\def \mQ {\mathcal Q}
\def \mR {\mathcal R}
\def \mS {\mathcal S}

\def \mU {\mathcal U}
\def \mV {\mathcal V}

\def \bX {{\bar X}}
\def \bh {{\bar h}}
\def \bc {{\bar c}}

\def \lag {\langle}
\def \rag {\rangle}

\def \p {\partial}
\def \f {\frac}

\def \nn {\nonumber}

\def \mc {\mathcal}

\def \eps {\epsilon}
\def \s {\sigma}
\def \lam {\lambda}

\def \lt {\left}
\def \rt {\right}

\def \sr {\sqrt}
\def \td {\tilde}
\def \hs {\hspace}

\def \inf {\infty}

\def \r {\rho}
\def \tr {\mathrm{tr}}
\def \Tr {{\textrm{Tr}}}

\begin{document}

\title{\textbf{Holographic R\'enyi entropy in  AdS$_3$/LCFT$_2$ correspondence}}
\author{
Bin Chen$^{1,2,3,4}$\footnote{bchen01@pku.edu.cn},\,
Feng-yan Song$^{1}$\footnote{songfy@pku.edu.cn}
and
Jia-ju Zhang$^{1}$\footnote{jjzhang@pku.edu.cn}
}
\date{}
\maketitle

\begin{center}
{\it
$^{1}$Department of Physics and State Key Laboratory of Nuclear Physics and Technology,\\Peking University, 5 Yiheyuan Rd, Beijing~100871, P.~R.~China\\
$^{2}$Collaborative Innovation Center of Quantum Matter, 5 Yiheyuan Rd, Beijing~100871, P.~R.~China\\
$^{3}$Center for High Energy Physics, Peking University, 5 Yiheyuan Rd, Beijing~100871, P.~R.~China\\
$^{4}$Beijing Center for Mathematics and Information Interdisciplinary Sciences,\\105 W~3rd~Ring Rd N, Beijing~100048, P.~R.~China
}
\vspace{10mm}
\end{center}

\begin{abstract}

  The recent study  in AdS$_3$/CFT$_2$ correspondence shows that the tree level contribution and 1-loop correction of holographic R\'enyi entanglement entropy (HRE) exactly match the direct CFT computation in the large central charge limit. This allows the R\'enyi entanglement entropy to be a new window to study the AdS/CFT correspondence. In this paper we generalize the study of R\'enyi entanglement entropy in pure AdS$_3$ gravity to the  massive gravity theories at the critical points. For the cosmological topological massive gravity (CTMG),  the dual conformal field theory (CFT) could be a chiral conformal field theory or a logarithmic conformal field theory (LCFT), depending on the asymptotic boundary conditions imposed. In both cases, by studying the short interval expansion of the R\'enyi entanglement entropy of two disjoint intervals with small cross ratio $x$, we find that the classical and 1-loop HRE  are in exact match with the CFT results, up to order $x^6$. To this order, the difference between the massless graviton and logarithmic mode can be seen clearly. Moreover, for the cosmological new massive gravity (CNMG) at critical point, which could be dual to a logarithmic CFT as well, we find the similar agreement in the CNMG/LCFT correspondence. Furthermore we read the 2-loop correction of graviton and logarithmic mode to HRE from CFT computation. It has distinct feature from the one in pure AdS$_3$ gravity.

\end{abstract}

\baselineskip 18pt
\thispagestyle{empty}

\newpage

\tableofcontents
\section{Introduction}

Entanglement entropy has been under active study in the past decade. It is defined as the von Neumann entropy of reduced density matrix $\r_A$ of a subsystem A
\be
S_A=-\Tr_A \r_A\log \r_A.
\ee
For a quantum field theory, the entanglement entropy is called the geometric entropy as the leading contribution is proportional to the area of the boundary of the subsystem\cite{Bombelli:1986rw,Srednicki:1993im}. More generally, with the reduced density matrix, one may define the R\'enyi entanglement entropy  as
\be
S_A^{(n)}=-\f{1}{n-1} \log \Tr_A \r_A^n.
\ee
It is easy to see that the entanglement entropy and the R\'enyi entropy are related by
\be
S_A=\lim_{n \to 1} S_A^{(n)}.
\ee
One can also define the R\'enyi mutual information of two subsystems $A$ and $B$ as
\be
I_{A,B}^{(n)}=S_{A}^{(n)}+S_{B}^{(n)}-S_{A\cup B}^{(n)}.
\ee
From its definition, the R\'enyi entropy could be calculated via the replica trick\cite{Callan:1994py}. However, this trick leads to the computation of the partition function on a spacetime manifold with nontrivial topology. For example, in two-dimensional quantum field theory on complex plane, the $n$-th R\'enyi entropy of $N$ intervals requires a partition function on a Riemann surface of genus $(n-1)(N-1)$. Therefore even though the R\'enyi entropy is easier  than the entanglement entropy, it is still quite hard to compute  in practice.

The AdS/CFT correspondence provides an effective tool to study the entanglement and the R\'enyi entropies. It was firstly proposed by Ryu and Takayanagi\cite{Ryu:2006bv,Ryu:2006ef} that  the entanglement entropy of subregion $A$ in a conformal field theory (CFT) could be holographically given by the area of a minimal surface which is homogeneous to $A$ in the dual AdS gravity. This so-called holographic entanglement entropy has been studied intensely since its proposal, see good reviews \cite{Nishioka:2009un,Takayanagi:2012kg} for complete references. For the R\'enyi entropy, it could be calculated holographically in a similar way\cite{Headrick:2010zt,Hung:2011nu}. Very recently the RT formula or prescription has been proved from various points of view. In \cite{Hartman:2013mia,Faulkner:2013yia} the RT formula has been proved in AdS$_3$/CFT$_2$ case. For general case, the RT formula has been shown to be true from the point of view of the generalized gravitational entropy \cite{Lewkowycz:2013nqa}\footnote{For earlier efforts to prove RT formula, see for examples \cite{Fursaev:2006ih,Casini:2011kv}.}.
One essential point in these proofs is to find the gravitational configurations in applying the replica trick. This turns out to be a subtle issue and has not been well-understood in general cases. However, in AdS$_3$/CFT$_2$ case, the bulk gravitational configurations ending on a higher-genus Riemann surface can be constructed explicitly without trouble.
Then the Euclidean gravity action of the configuration gives the leading contribution to the holographic R\'enyi entropy (HRE). Moreover, with the gravitational configuration, the gravitational 1-loop correction has been considered in \cite{Barrella:2013wja}. This 1-loop quantum correction is essential to the mutual information \cite{Faulkner:2013ana}.
From AdS$_3$/CFT$_2$ correspondence for pure AdS gravity, the central charge $c=\f{3l}{2G}$ is inversely proportional to the Newton coupling constant $G$, so the large central charge $c$ limit  corresponds to the weak coupling limit in the gravity. Therefore the classical, quantum 1-loop, 2-loop, ... contributions to HRE correspond to the CFT contributions proportional to $c$, $c^0$, $\f{1}{c}$, ..., respectively \cite{Headrick:2010zt}.
It is remarkable that for the two-interval case with a small cross ratio $x$, the  classical and 1-loop contributions to the HRE are in exact agreement with the CFT results up to order $x^8$ \cite{Chen:2013kpa,Chen:2013dxa}. These facts provide nontrivial support of the holographic computations of the R\'enyi entropy, not only at the classical level but also at quantum 1-loop level.

On the other hand, the R\'enyi entropy could be taken as a new window to study the AdS$_3$/CFT$_2$ correspondence. For the two-interval case, the second R\'enyi entropy is actually the torus partition function, which is usually the first check on the possible correspondence. The higher rank R\'enyi entropy is in general hard to compute, but could be calculated order by order by using the operator product expansion of the twist operators in the small interval limit \cite{Headrick:2010zt,Calabrese:2010he,Chen:2013kpa,Chen:2013dxa,Perlmutter:2013paa}. In \cite{Chen:2013kpa}, we considered the pure AdS$_3$ gravity and the vacuum Verma module of its CFT dual, and found exact agreements. Recently such investigation has been generalized to the holographic R\'enyi entropy for Higher spin gravity/CFT with $W$ symmetry correspondence in \cite{Chen:2013dxa,Perlmutter:2013paa}. In this paper, we would like to extend our study to the topologically massive gravity with negative cosmological constant (CTMG) \cite{Deser:1981wh,Deser:1982vy} and cosmologically new massive gravity (CNMG) \cite{Bergshoeff:2009hq,Bergshoeff:2009aq}, both of which at critical points have been conjectured to be dual to some kinds of CFT under appropriate asymptotic boundary conditions.

In three dimensional (3D) topologically massive gravity (TMG) with a negative cosmological constant, there is a gravitational Chern-Simons term in the action. Off the critical point, there could be a massive fluctuation around the AdS$_3$ vacuum, besides two massless modes. It turns out that off the critical point the theory is ill-defined due to the negative energy of the massive mode. At the critical points, the theory could be well-defined but becomes chiral after imposing Brown-Henneaux boundary conditions, as the local massive mode becomes degenerate with the left massless mode and the only degree of freedom is a massless boundary graviton. It was conjectured that the chiral gravity is dual to a chiral CFT with only right-mover \cite{Li:2008dq,Strominger:2008dp}. However, even at the critical point, there is actually a logarithmic mode \cite{Carlip:2008jk,Carlip:2008eq,Grumiller:2008qz}, if one does not impose the Brown-Henneaux boundary conditions. Moreover it has been found that there exist another set of consistent boundary conditions to include the logarithmic mode, and the resulting quantum gravity is proposed to be dual to a logarithmic CFT (LCFT) \cite{Grumiller:2008es,Henneaux:2009pw,Maloney:2009ck,Grumiller:2009mw,Gaberdiel:2010xv}. The lesson is that the quantum gravity is defined with respect to the asymptotic boundary conditions.

Another interesting class of 3D massive gravity is the so-called new massive gravity (NMG). Due to the presence of the higher derivative terms, the theory generically has massive gravitons, similar to CTMG. In this work, we focus on the AdS vacuum, in which NMG is called cosmological NMG (CNMG). For CNMG, there exists a critical point where the massive modes disappear and the logarithmic modes appear \cite{Liu:2009bk,Bergshoeff:2009aq}. Very interestingly, similar to CTMG, there are more than one set of consistent boundary conditions to define the quantum gravity. In general, one may impose the usual Brown-Henneaux boundary conditions to set up a AdS$_3$/CFT$_2$ correspondence,  but at critical point there could be other sets of consistent boundary conditions to include the logarithmic mode(s)\cite{Liu:2009kc}. Therefore at the critical point, there is a CNMG/LCFT correspondence \cite{Grumiller:2009sn,Gaberdiel:2010xv}, even though the central charges in this case are vanishing.


In this work, we study the R\'enyi entropy of two disjoint intervals in the framework of CTMG/CFT  correspondence. We show that for a general gravity theory with a AdS$_3$ vacuum, the classical HRE is similar to the one in pure AdS gravity
\be
\f{\p S_n}{\p z_i}=-\f{n(c_L+c_R)}{12}\g_i,
\ee
with the only difference on the sum of the central charges of the dual CFT. For the CTMG, as the sum of the central charges is the same as the pure AdS gravity, so is the classical HRE. However we find that the 1-loop quantum corrections depend on the choice of asymptotic boundary conditions. On the other side, we compute the OPE of twist operators in the small interval limit in the corresponding CFT duals. For the chiral CFT, the computation is relatively easy, but for the logarithmic CFT, we have to treat it carefully. In both the chiral and logarithmic cases, we find that HRE and CFT results are in exact agreement up to order $x^6$. We furthermore discuss the holographic R\'enyi entropy in CNMG at the critical point. In this case, the classical contribution to the entropy is simply vanishing as the central charges are zero, but the quantum corrections are not vanishing. We find agreement in the CNMG/LCFT correspondence as well. Our results support the correspondence between the three-dimensional critical massive gravity and two-dimensional logarithmic CFT.

The remaining of the paper is arranged as follows. In Section~\ref{s2} we review the fundamental facts on the chiral gravity and log gravity in the context of CTMG. Then we compute the holographic R\'enyi entropy in CTMG and CNMG at critical points. In Section~\ref{s3} after introducing the logarithmic CFT and showing how to study the OPE of twist operators in it,  we compute the R\'enyi entropy in the small interval limit carefully.  In Section~\ref{s4}, we end with conclusion and discussion. We put some summation formulas into Appendix~\ref{sa}.

{\bf Note added} While we are finishing the manuscript, there appeared a paper \cite{Perlmutter:2013paa}, which has some overlaps with Section~\ref{s2} of the present work.

\section{Holographic R\'enyi entropy in critical massive gravity} \label{s2}

We are going to compute  the holographic R\'enyi entropy in cosmological massive gravity at the critical point for two intervals with small cross ratio in CFT. As we are considering the entanglement entropy in the vacuum state of CFT, we focus on the AdS$_3$ vacuum. The classical solutions found in pure AdS$_3$ gravity are always the solutions of these theories. Thus the gravitational configurations corresponding to the $n$-th R\'enyi entropy are the same as the ones worked out in \cite{Faulkner:2013yia,Barrella:2013wja}. They are just the quotient of AdS$_3$ by the Schottky group. Locally they are diffeomorphic to global AdS$_3$, or in other words they satisfy the relations $R_{\m\n}=\f{1}{3}g_{\m\n}R=-\f{2}{l^2}g_{\m\n}$ with $l$ being the AdS radius.

The tree level contribution comes from the Euclidean action of the configurations, including the boundary terms. Here we may consider a quite general 3D gravity theory with a AdS$_3$ vacuum. Let us start from the Euclidean action
\be
I=\f{1}{16\pi G}\int d^3x \sr{g} {\mathcal L}(g_{\m\n},\nabla_\m, R_{\m\n})+ I_{{bndy}},
\ee
and forget about the gravitational Chern-Simons terms for a while. The boundary term $I_{{bndy}}$ is not essential in the following discussion. The AdS$_3$ vacuum is of the radius $l$, which should be determined from the equation of motion of the theory. In Euclidean gravity, the vacuum becomes a hyperbolic space $H$, whose metric could be written in terms of Poincar\'e coordinates as
\be
ds^2=\f{l^2}{z^2}(dx^2+dy^2+dz^2).
\ee
The boundary of $H$ is a Riemann sphere at $z=0$. The boundary of the above mentioned gravitational configurations are Riemann surfaces, which could be taken as the quotients of the Riemann sphere by the Schottky groups.

The bulk action for any  gravitational configuration above is
\be
I_{bulk}=\f{{\mathcal L}_m}{16\pi G}\int d^3x \sqrt{g},
\ee
where ${\mathcal L}_m$ is the value of the Lagrangian density at the AdS vacuum, and is a constant. This action is obviously divergent and needs regularization. The standard way is to introduce a plane at $z=\epsilon$. Then the quadratic divergence could be cancelled by the boundary action, i.e.
\bea
&& I_\e=\f{{\mathcal L}_m}{16\pi G}(V_\e-\f{l}{2}A_\e)\nn\\
&& \phantom{I_\e}
       =\f{{\mathcal L}_ml^3}{16 G}(V_{reg}-(2g-2)\ln\e),
\eea
where $g$ is the genus of the Riemann surface at the boundary.
The logarithmic term could not be canceled by a local counter term. It is actually related to the Weyl anomaly of a Riemann surface of genus $g$ \cite{Krasnov:2000zq}. This allows us to determine the central charge of the dual CFT
\be
c=\f{{3\mathcal L}_ml^3}{8 G},
\ee
which has been found from another point of view in \cite{Kraus:2005vz}.
More interestingly, it turns out that the regularized action $I_\e$ is related to the Liouville action of the Riemann surface\cite{Zograf:1988}. A careful study along the line in \cite{Faulkner:2013yia} shows that the regularized action could be expressed in terms of the accessory parameters characterizing the Schottky uniformization,
\be
\f{\p S_n}{\p z_i}=-\f{cn}{6}\g_i,
\ee
where $\g_i$ is fixed by the monodromy problem of an ordinary differential equation. Therefore for any 3D gravity with a AdS$_3$ vacuum, the classical HRE could be determined by this formula. The only difference is on the central charge.

If the  theory includes the gravitational Chern-Simons term, it does not change the above argument. The presence of the CS term may induce diffeomorphism anomaly in the stress tensor, but the sum of the central charges are invariant\cite{Kraus:2005zm,Solodukhin:2005ah}. As a result, the above relation should be modified a little bit
\be
\f{\p S_n}{\p z_i}=-\f{n(c_L+c_R)}{12}\g_i.
\ee
Note that for a Minkowski 2D CFT  the left- and right-moving central charges are denoted by $c_{L,R}$, which  are just the holomorphic and antiholomorphic central charges $c, \bar c$ for the Euclidean version of the CFT.  For the case of  two intervals with small cross ratio $x$, one can  get the classical part of the holographic R\'enyi mutual information to order $x^6$\cite{Hartman:2013mia,Faulkner:2013yia,Barrella:2013wja}
\bea \label{incl}
&& I_n^{cl}=\frac{(c+\bar c) (n-1) (n+1)^2 x^2}{288 n^3}+\frac{(c+\bar c)(n-1)(n+1)^2 x^3}{288 n^3}  \nn\\
&& \phantom{I_n^{cl}=}
              +\frac{(c+\bar c) (n-1) (n+1)^2 \left(1309 n^4-2 n^2-11\right) x^4}{414720 n^7}  \nn\\
&& \phantom{I_n^{cl}=}
              +\frac{(c+\bar c) (n-1) (n+1)^2 \left(589 n^4-2 n^2-11\right) x^5}{207360 n^7}   \\
&& \phantom{I_n^{cl}=}
              +\frac{(c+\bar c) (n-1) (n+1)^2 \left(805139 n^8-4244 n^6-23397 n^4-86 n^2+188\right) x^6}{313528320 n^{11}}+O(x^7). \nn
\eea
For CTMG theory, the sum of the central charges are the same as the pure AdS$_3$ vacuum. Thus the classical HRE is the same as the one in pure AdS$_3$ gravity.

Note that the above discussion has nothing to do with the choice of the asymptotic boundary conditions. And the result is true for any AdS$_3$ vacuum, not restricted to the critical points. Therefore the classical HRE takes a universal form, depending only on the central charges.


For the 1-loop correction to the HRE, the situation is more complex. We have to consider all the possible fluctuations around the gravitational configurations. As these configurations are locally diffeomorphic to AdS$_3$, we can study the linearized equation around the AdS vacuum to read the fluctuations. Another closely related subtle issue is the imposing of asymptotic boundary conditions, since different boundary conditions define different quantum gravities and their CFT duals. In this section, we discuss two kinds of massive gravity theories at the critical points and the 1-loop correction to the holographic R\'enyi entropy in these theories.

\subsection{CTMG at critical point}

The topologically massive gravity (TMG) has been studied for a long time. From the view of AdS/CFT, the CFT dual to CTMG has different central charges on the left and right sector due to the existence of diffeomorphism anomaly.  The action of CTMG is
\be \label{TMGaction} S = \f{1}{16\pi G}\int d^3x \sqrt{-g} \left[R+\f{2}{l^2}+\f{1}{\mu}\eps^{\lambda\mu\nu}\G^{\r}_{\s\lam}(\p_{\mu}\G^{\s}_{\r\nu}+\f{2}{3}\G^{\s}_{\kappa\mu}\G^{\kappa}_{\r\nu})\right].\ee
Here $l$ is AdS radius and $G$ is the Newton constant. We follow the convention that $G$ is positive and $\m l\geq 1$ such that the energy of the BTZ black hole is positive and the central charges are always positive.
The AdS$_3$ spacetime is the solution of CTMG
\be
ds^2={l^2}\Big( -\lt(dx^+\rt)^2-\lt(dx^-\rt)^2 -2\cosh (2\r) dx^+dx^- + d\r^2 \Big).
\ee
The linear fluctuations around the AdS$_3$ vacuum obey a third order differential equation. As a result if $\m l\neq 1$, there are two massless boundary gravitons $h^L, h^R$ and a local massive graviton $h^M$. When $\m l=1$, the local massive mode becomes degenerate with the left massless mode so that the only degree of freedom could be the right massless boundary graviton $h^R$. However, there is actually a new mode at critical point $\m l=1$, defined by
 \be
h^{log} = \lim_{\mu l \rightarrow 1}\f{h^{M}-h^{L}}{\mu l -1}.\ee
This mode has component that increases linearly when $\r$ goes to infinity, so it is called logarithmic mode \cite{Grumiller:2008qz}. It satisfies the linearized equation as well, and moreover carries negative energy.


Quantum gravity in AdS$_3$ is defined with respect to the asymptotic boundary conditions. For CTMG, one may impose  the Brown-Henneaux asymptotically
boundary conditions \cite{Brown:1986nw}
\be
\left(
\begin{array}{ccc}
 h_{++} = {O}(1) & h_{+-} = {O}(1)  & h_{+\r} = {O}(e^{-2\r}) \\
 h_{-+} = h_{+-} & h_{--} = {O}(1) & h_{-\r} = {O}(e^{-2\r})  \\
 h_{\r+} = h_{+\r} & h_{\r-} = h_{-\r} & h_{\r\r} = {O}(e^{-2\r}) \label{BHcondition}
\end{array}
\right).
\ee
Accordingly, the generator of asymptotic symmetry group (ASG) is \cite{Strominger:2008dp}
\be
Q \sim \left(1+\f{1}{\mu l}\right)\int_{\p \Sigma}dx^{+}h_{++}\eps^{+}+\left(1-\f{1}{\mu l}\right)\int_{\p \Sigma}dx^{-}h_{--}\eps^{-}.
\ee
This set of boundary conditions is well-defined for generic value of $\m l$, and leads to two copies of Virasoro algebra with the central charges
\be
c_L=\left(1-\f{1}{\mu l}\right)\f{3l}{2G}, \hs{3ex}c_R=\left(1+\f{1}{\mu l}\right)\f{3l}{2G}.\ee
At the critical point $\mu l = 1$,
the generator becomes
\be
Q \sim \int_{\p \Sigma}dx^{+}h_{++}\eps^{+}.
\ee
There is only one set of Virasoro algebra, corresponding to the right-moving sector of a CFT. Note that the Brown-Henneaux boundary conditions exclude the logarithmic mode.
 Therefore the CTMG theory becomes a chiral gravity at the critical point. It could be holographically dual to a chiral CFT with pure right sector\cite{Li:2008dq}
 \be  c_{L} = 0, ~ c_R = \f{3l}{G}.  \label{centralChiral}\ee

On the other hand, one can impose another set of consistent boundary conditions to include the logarithmic mode\cite{Grumiller:2008es,Maloney:2009ck}. To make the right-moving charge be well defined, the boundary conditions should be
\be
\left(
\begin{array}{ccc}
 h_{++} = {O}(1) & h_{+-} = {O}(1)  & h_{+\r} = {O}(e^{-2\r}) \\
 h_{-+} = h_{+-} & h_{--} = {O}(\r) & h_{-\r} = {O}(\r e^{-2\r})  \\
 h_{\r+} = h_{+\r} & h_{\r-} = h_{-\r} & h_{\r\r} = {O}(e^{-2\r)}
\end{array}
\right),\label{logboundary}
\ee
under which the left- and right-moving charges $Q^L$ and $Q^R$ are both finite.
Even though the logarithmic mode carries negative energy, the CTMG including such mode is conjectured to be dual a logarithmic CFT with the same central charges (\ref{centralChiral}).

In the  log gravity, using the method of heat kernel \cite{Giombi:2008vd,David:2009xg}, one can read the 1-loop partition function \cite{Gaberdiel:2010xv}
\be Z_{TMG}^{1-loop}= \prod_{r=2}^{\infty}\f{1}{|1-q^r|^2}\prod_{m=2}^{\infty}\prod_{\bar{m} = 0}^{\infty}\f{1}{1-q^m\bar{q}^{\bar{m}}} \label{partitionTMG}. \ee
The first part being the product over $r$ is the contribution from massless boundary gravitons, while the remaining part is the contribution from the logarithmic mode.
According to \cite{Barrella:2013wja}, the on-shell 1-loop partition function could be
\be \label{hartnollsmethod}
\log{Z_{TMG}^{1-loop}} = - \sum_{\g\in P} \sum_{r=2}^{\infty}\log{(|1-q_{\g}^r|)}
                         - \f{1}{2}\sum_{\g\in P}\sum_{m=2}^{\infty}\sum_{\bar{m} = 0}^{\infty}\log{(1-q_{\g}^m \bar q_{\g}^{\bar{m}})}.
\ee
Here $P$ is a set of representatives of primitive conjugacy classes of the Schottky group $\G$. And the $q_{\g}$ is one of the two eigenvalues of the matrix $\g \in \G$ with the condition that $|q_{\g}|<1$. We consider the case of two short intervals on complex plane with small cross ratio. Using the conformal invariance, we can map the subsystem to $A=(-\inf,-1]\cup[-y,y]\cup[1,+\inf)$ as in \cite{Barrella:2013wja}, and the cross-ratio will be
\be x = \f{4y}{(1+y)^2}.\ee
To the order of $x^6$, we only need the so-called consecutively decreasing words (CDWs)
\be \g_{k,m} = M_2^{m+k}T^{-1}M_2^{k}TM_2^{-m} .\ee
The 1-loop result in the gravity side is separable. For the massless boundary gravitons there is the 1-loop R\'enyi mutual information \cite{Barrella:2013wja}
\bea \label{ingr}
&& I_{n, GR}^{1-loop}=\frac{(n+1) \left(n^2+11\right) \left(3 n^4+10 n^2+227\right) x^4}{3628800 n^7} \nn\\
&& \phantom{I_n^{1-loop}=}
+\frac{(n+1) \left(109 n^8+1495 n^6+11307 n^4+81905 n^2-8416\right) x^5}{59875200 n^9}\nn\\
&& \phantom{I_n^{1-loop}=}
+\frac{(n+1) x^6}{523069747200 n^{11}}\left(1444050 n^{10}+19112974 n^8+140565305 n^6+1000527837 n^4 \rt. \nn\\
&& \phantom{I_n^{1-loop}=} \lt.  -167731255 n^2-14142911\right) +O\left(x^7\right),
\eea
and the 1-loop contribution from the log mode is
\bea \label{inlog}
&& I_{n, log}^{1-loop}= \frac{(n+1) \left(n^2+11\right) \left(3 n^4+10 n^2+227\right) x^4}{7257600 n^7} \nn\\
&& \phantom{I_n^{1-loop}=}+\frac{(n+1) \left(109 n^8+1495 n^6+11307 n^4+81905 n^2-8416\right) x^5}{119750400 n^9} \nn\\
&& \phantom{I_n^{1-loop}=}+\frac{(n+1) x^6}{2615348736000 n^{11}}\left(3610816 n^{10}+47796776 n^8+351567243 n^6+2502467423 n^4 \rt. \nn\\
&& \phantom{I_n^{1-loop}=} \lt. -412426559 n^2+10856301\right)+O\left(x^7\right).
\eea
Then the 1-loop R\'enyi mutual information for the TMG as log gravity is
\be \label{intmg}
I_{n, TMG}^{1-loop}=I_{n, GR}^{1-loop}+I_{n, log}^{1-loop}.
\ee

It is remarkable that the 1-loop contributions from the massless graviton and the logarithmic mode are the same at the first two leading orders, i.e. at order $x^4$ and $x^5$. The difference between these two  modes appears first at order $x^6$. This requires us to study the quasiprimary operators to level six, in order to see this difference in CFT.

For the chiral gravity, the computation is relatively easier. By imposing the  Brown-Henneaux boundary conditions, the logarithmic mode should be truncated such that
the 1-loop partition function of the chiral gravity takes the form \cite{Maloney:2009ck}
\be
Z_{chiral}^{1-loop} = \prod_{r=2}^{\infty}\f{1}{1-q^r}. \label{chiralpartition}
\ee
Note that this is different from the result in pure AdS$_3$ gravity since due to the enhanced gauge symmetry there is only one massless graviton in the chiral gravity.
For the 1-loop HRE, we get half of the result in \cite{Barrella:2013wja}, i.e. half of (\ref{ingr}).

\subsection{Critical CNMG}

The action of NMG can be written as
\be S=\f{1}{16\pi G}\int dx^3 \sqrt{-g}[\sigma R+\f{1}{m^2}(R_{\mu \nu}R^{\mu \nu}-\f{3}{8}R^2)-2\lambda m^2],\ee
where $\lambda$ is a dimensionless cosmological constant, $\sigma$ is the sign of the Einstein-Hillbert term and $m$ is the mass parameter. Up to the values of the parameters, there are various vacua in NMG\cite{Bergshoeff:2009aq}. Here we focus on the so-called CNMG with AdS$_3$ as a vacuum.
The AdS radius is determined by the real solution of $1/l^2 = 2m^2(\sigma \pm \sqrt{1+\lambda})$.

In general, there are two massless boundary gravitons $h^L, h^R$, and two local massive gravitons $h^{m\pm}$. At the critical point \cite{Liu:2009bk}
\be
2m^2l^2=-\s,
\ee
the massive modes $h^{m\pm}$ coincide with the massless modes $h^L$ and $h^R$,  and there appear the left- and right-moving logarithmic modes $h^{log}_L$ and $h^{log}_R$ \cite{Liu:2009kc,Grumiller:2009sn}.

For CNMG, the Brown-Henneaux boundary conditions (\ref{BHcondition}) are always well-defined. The corresponding ASG has the central charges
\be
c_L = c_R = \f{3l}{2G_N}(\sigma + \f{1}{2l^2m^2}).
\ee
At the critical point $2m^2l^2=-\s$, both the central charges of CNMG are zero \cite{Liu:2009bk}. Moreover, similar to the CTMG case,  there
exists another set of consistent boundary conditions at the critical point to
include both the left- and right-moving logarithmic modes
\be
\left(
\begin{array}{ccc}
 h_{++} = {O}(\r) & h_{+-} = {O}(1)  & h_{+\r} = {O}(\r e^{-2\r}) \\
 h_{-+} = h_{+-} & h_{--} = {O}(\r) & h_{-\r} = {O}(\r e^{-2\r})  \\
 h_{\r+} = h_{+\r} & h_{\r-} = h_{-\r} & h_{\r\r} = {O}(e^{-2\r})
\end{array}
\right),\label{CNMGlogboundary}
\ee
Actually, as discussed carefully in \cite{Liu:2009kc}, there are two other sets of consistent boundary conditions, which include left or right logarithmic mode respectively. In all these cases, the conserved charges are well-defined, though the central charges are vanishing. In this work, we are interested in the CNMG with both of the logarithmic modes.

The holographic R\'enyi entropy in the CNMG is subtle due to the presence of higher derivative terms in the action. In \cite{Bhattacharyya:2013gra}, the holographic entangle entropy for one interval has been discussed. The tree level contribution is proportional to the central charge of possible dual CFT, in match with the CFT prediction. This is true for general $n$-th R\'enyi entropy from our discussion at the beginning of this section. As the central charges of the CNMG at the critical point is zero, the classical HRE is zero accordingly. However, the 1-loop correction is nontrivial as there are various fluctuations around the configurations.

It is straightforward to calculate the 1-loop R\'{e}nyi entropy of CNMG at the critical point. As the 1-loop partition function of CNMG is \cite{Gaberdiel:2010xv}
\be Z_{NMG}^{1-loop} = \prod_{r=2}^{\infty}\f{1}{|1-q^r|^2}\prod_{m=2}^{\infty}\prod_{\bar{m} = 0}^{\infty}\f{1}{1-q^m\bar{q}^{\bar{m}}}\prod_{p=0}^{\infty}\prod_{\bar{p} = 2}^{\infty}\f{1}{1-q^p\bar{q}^{\bar{p}}},\ee
and then the 1-loop R\'enyi mutual information is
\be \label{innmg}
I_{n, NMG}^{1-loop}=I_{n, GR}^{1-loop}+2I_{n, log}^{1-loop},
\ee
with $I_{n, GR}^{1-loop}$ and $I_{n, log}^{1-loop}$ being the same as (\ref{ingr}) and (\ref{inlog}) respectively.

\section{R\'enyi entropy in LCFT}\label{s3}

In this section, we investigate the computation of the R\'enyi entropy in the logarithmic  conformal field theory (LCFT). We first show how to realize the $c=0$ LCFT by introducing extra primary field into an ordinary CFT and taking $c\to 0$ limit. Accordingly we discuss how to construct the quasiprimary operators in this realization. After introducing the OPE of twist operators in the small interval limit in LCFT, we calculate the R\'enyi mutual information of two disjoint intervals with a small cross ratio $x$ to order $x^6$ and find agreement with the classical and 1-loop HRE in bulk gravity theories. Moreover, we read the 2-loop correction to HRE, which shows some novel features.

\subsection{Basics of LCFT}

First let us review briefly the basics of a LCFT. In normal conformal field theory, the primary operators and their descendants form a complete set. Every operator can be expressed as a linear combination of this set. The states in the Verma module are eigenstates of $L_0$ and the correlation functions do not have logarithmic terms. This means that the matrix of $L_0$ is diagonal if we use the states in the Verma modules as basis. However, in some non-unitary theories, the four-point functions of some operators do have logarithmic terms. This indicates  the existence of additional operators, which together with the ordinary ones, form the basis of the $L_{0}$. The matrix of $ L_0$ under this basis is not diagonal, but performs like a Jordan cell. In this case, the primary operators and their descendants cannot be a complete set, and one must add the contribution of the pseudo-operators to form a complete set. The correlation functions involving the pseudo-operators then have the logarithmic terms \cite{Gurarie:1993xq}. There are various kinds of LCFT as classified in \cite{Kogan:2001ku,Kogan:2002mg}, but here we only discuss the type with vanishing central charge and nondegenerate vacuum.

The usual way to study LCFT is just adding a nilpotent part to the conformal weight of the primary field \cite{MoghimiAraghi:2000qn,Flohr:1997wm}. One can use this nilpotent formalism to rewrite the correlation functions in the LCFT. However, there is another approach to describe the LCFT when the central charge $c = 0$. It could be taken as the limit  of an ordinary CFT with varying  central charge $c$ \cite{Cardy:1999,Gurarie:1999,Cardy:2001,Kogan:2001ku,Kogan:2002mg,Grumiller:2013at,Cardy:2013rqg}. In the following we will use the convention in \cite{Kogan:2002mg}.

The LCFT dual to the log gravity in CTMG has vanishing holomorphic central charge $c=0$ but nonvanishing antiholomorphic central charge $\bar c\neq 0$, and only its holomorphic part is logarithmic. It could be viewed as the $c \to 0$ limit of a normal CFT. The normal CFT has the stress tensor $T(z)$ and $\bar T(\bar z)$ with central charges $(c,\bar c)$, as well as a primary operator $X(z,\bar z)$ with conformal weights $(h,\bar h)=(2+\e(c),\e(c))$. The scaling dimension $\D$ and the spin $s$ of the primary field are respectively
\be
\D=h+\bar h=2+2\e(c), ~~~ s=h-\bar h=2.
\ee
There is the relation
\be \e(c)=-\f{1}{b}c+O(c^2), \ee
in which the constant $b$ is called the new anomaly. For the LCFT dual to the log gravity there is \cite{Grumiller:2009mw}
\be
b=-\f{3l}{G}.
\ee
The operator $X$ could be normalized such that
\be
\lag X(z_1,\bar z_1)X(z_2,\bar z_2) \rag_C=\f{\a_X}{z_{12}^{2h}\bar z_{12}^{2\bar h}},
\ee
where $z_{ij}\equiv z_i-z_j$, and
\be
\a_X=\f{B(c)}{c}, ~~~ B(c)=-\f{1}{2}+B_1 c +O(c^2).
\ee
Note that the monodromy of the two-point function requires that $s$ must be an integer or a half integer.
As usual there are two-point functions
\bea
&& \langle T(z_1)T(z_2)\rangle_C   =  \f{c}{2z_{12}^4}, \nn\\
&& \langle T(z_1)X(z_2,\bar z_2)\rangle_C = 0.
\eea
The logarithmic partner of $T(z)$ is defined as
\be\label{t}
t(z,\bar z)=\f{b}{c} T(z)+b X(z,\bar z).
\ee
Then one can take the limit $c\to 0$ and get the two-point funtions of the LCFT \cite{Kogan:2002mg}
\bea
&& \lag T(z_1)T(z_2) \rag =0, \\
&& \lag T(z_1) t(z_2,\bar{z_2})\rag = \f{b}{2z_{12}^4},\\
&& \lag t(z_1,\bar{z_1})t(z_2,\bar{z_2}) \rag = \f{B_1-b\ln{(|z_{12}|^2)}}{z_{12}^4}.
\eea
Note that the constant $B_1$ can be set to zero with a redefinition of $t$, and we will adopt $B_1=0$ hereafter.
In the calculation below we will also need the three-point functions \cite{Kogan:2002mg}
\bea
&& \lag T(z_1)X(z_2,\bar z_2)X(z_3,\bar z_3) \rag_C=\f{h\a_X}{z_{12}^2 z_{13}^2 z_{23}^{2h-2} \bar z_{23}^{2\bar h}}, \nn\\
&& \lag \bar T(\bar z_1)X(z_2,\bar z_2)X(z_3,\bar z_3) \rag_C=\f{\bar h\a_X}{z_{23}^{2 h} \bar z_{12}^2 \bar z_{13}^2 \bar z_{23}^{2 \bar h-2}}, \nn\\
&& \lag X(z_1,\bar z_1)X(z_2,\bar z_2)X(z_3,\bar z_3) \rag_C=\f{C_{XXX}}{\lt( z_{12}z_{13}z_{23} \rt)^h \lt( \bar z_{12}\bar z_{13}\bar z_{23} \rt)^{\bar h}}.
\eea
The structure constant is
\be
C_{XXX}=\f{D(c)}{c^2}, ~~~ D(c)=2-\f{3}{2b}c+O(c^2).
\ee
Note that only when $s$ is an even integer can we have $C_{XXX}$ nonvanishing.

The LCFT dual to the critical NMG is a little different to the one dual to the log gravity. In this case, both the holomorphic and antiholomorphic central charges of the LCFT are zero. And now the new anomaly is \cite{Grumiller:2009sn}
\be
b=-\s \f{12l}{G}.
\ee
The LCFT can be viewed as the $c\to0$ limit of a normal CFT with equal holomorphic and antiholomorphic central charges $c=\bar c$. However, besides the primary operator $X(z,\bar{z})$ introduced above, another primary operator $\bar X(z,\bar z)$ with holomorphic and antiholomorphic conformal weights $(\bar h,h)=(\e(c),2+\e(c))$ has to be taken into account.  Note that the holomorphic conformal weight of $\bar X(z,\bar z)$ is exactly the antiholomorphic conformal  weight of $X(z,\bar z)$, and vice versa. Thus $\bX$ has the same scaling dimension but opposite spin to $X$.  For $\bar X$ we have the correlation functions
\bea
&& \lag \bar X(z_1,\bar z_1)\bar X(z_2,\bar z_2) \rag_C = \f{\a_\bX}{z_{12}^{2\bh}\bar z_{12}^{2h}},  \nn\\
&& \lag T(z_1)\bX(z_2,\bar z_2)\bX(z_3,\bar z_3) \rag_C = \f{\bh \a_\bX}{z_{12}^2 z_{13}^2 z_{23}^{2\bh-2} \bar z_{23}^{2h}}, \nn\\
&& \lag \bar T(\bar z_1)\bX(z_2,\bar z_2)\bX(z_3,\bar z_3) \rag_C = \f{h\a_\bX}{z_{23}^{2\bh} \bar z_{12}^2 \bar z_{13}^2 \bar z_{23}^{2  h-2}}, \nn\\
&& \lag \bX(z_1,\bar z_1)\bX(z_2,\bar z_2)\bX(z_3,\bar z_3) \rag_C=\f{C_{\bX\bX\bX}}{\lt( z_{12}z_{13}z_{23} \rt)^{\bh} \lt( \bar z_{12}\bar z_{13}\bar z_{23} \rt)^{h}},
\eea
where $\bX$ is normalized the same as $X$ such that $\a_\bX=\a_X$, and also  $C_{\bX\bX\bX}=C_{XXX}$. There are also the three-point functions
\bea
&& \lag X(z_1,\bar z_1)X(z_2,\bar z_2)\bX(z_3,\bar z_3) \rag_C
   =\f{C_{XX\bX}}{ z_{12}^{2h-\bh} z_{13}^\bh z_{23}^\bh \bar z_{12}^{2\bh-h} \bar z_{13}^h \bar z_{23}^h },  \nn\\
&& \lag \bar X(z_1,\bar z_1)\bar X(z_2,\bar z_2) X(z_3,\bar z_3) \rag_C
   =\f{C_{\bX\bX X}}{ z_{12}^{2\bh-h} z_{13}^ h z_{23}^ h \bar z_{12}^{2 h-\bh} \bar z_{13}^\bh \bar z_{23}^\bh },
\eea
with
\be
C_{XX\bX}=C_{\bX\bX X}=\f{F(c)}{c}, ~~~ F(c)=-\f{1}{2b}+O(c).
\ee
The logarithmic partner $t(z,\bar z)$ of $T(z)$ is defined the same as (\ref{t}), and the logarithmic partner $\bar t(z,\bar z)$ of $\bar T(\bar z)$ is defined similarly
\be\label{tb}
\bar t(z,\bar z)=\f{b}{c} \bar T(\bar z)+b \bar X(z,\bar z).
\ee

\subsection{Short interval expansion of R\'{e}nyi entropy}

In  2D CFT, the R\'{e}nyi entropy of single interval with length $\ell$ is proportional to its central charge \cite{Calabrese:2004eu,Calabrese:2009qy}
\be S_n = \f{c}{6}\lt( 1+\f{1}{n} \rt)\ln{\f{\ell}{\eps}}.\ee
with $\eps$ being the UV cutoff.
The computation of R\'{e}nyi entropy of $N$ intervals are much more complicated since we need to calculate the $2N$ point function of twist operators in a orbifold CFT. In this orbifold CFT, we need to make $n$ copies of the original CFT, so for simplicity we call this orbifold CFT as $CFT^n$. In this work, we focus our attention on the case of two disjoint intervals.

If the two intervals are far away, we can use the method of operator product expansion (OPE) of the twist operators at the ends of one interval \cite{Headrick:2010zt,Calabrese:2010he,Chen:2013kpa}. The nontrivial boundary conditions for $CFT^n$ could be replaced by the insertions of twist operators at the boundaries of all the intervals. The OPE of one twist operator and one antitwist operator can be written as
\be
\sigma(z,\bar{z})\td\sigma(0,0) = c_n\sum_{K}d_K\sum_{p,q\geq 0}\f{a_{K}^p}{p!}\f{\bar{a}_{K}^q}{q!}\f{1}{z^{2h_{\sigma} - h_K-p}z^{2\bar{h}_{\sigma} - \bar{h}_K-q}}\p^p\bar{\p}^q\Phi_{K}(0,0).
\ee
where $c_n$ is a constant number and  not important in our computation later. The summation $K$ is over all the independent quasiprimary operators of $CFT^n$.
The coefficients $a_{K}^p$ and $\bar{a}_{K}^q$ are respectively
\be a_{K}^p = \f{C_{h_K+p-1}^p}{C_{2h_K+p-1}^p},~\bar{a}_{K}^q = \f{C_{\bar{h}_K + q - 1}^q}{C_{2\bar{h}_K + q - 1}^q},\ee
with $C_{h_K+p-1}^p$, $\cdots$ being the binomial coefficients.
Both the twist and antitwist operators $\s$, $\td \s$ have the conformal weights \cite{Calabrese:2004eu,Calabrese:2009qy}
\be h_\s=\f{c}{24} \lt( n-\f{1}{n} \rt), ~~~ \bar h_\s=\f{\bar c}{24} \lt( n-\f{1}{n} \rt). \ee
The coefficients $d_K$ could be read from the one-point function of the corresponding quasiprimary operator \cite{Calabrese:2010he,Chen:2013kpa}
\be \label{dk}
d_{K} = \f{1}{\alpha_{K}\ell^{h_{K }+\bar{h}_{K}}}\lim_{z\rightarrow\inf}z^{2h_{K}}\bar{z}^{2\bar{h}_K}\langle\Phi_{K}(z,\bar{z}) \rangle_{\mc R_{n,1}} .
\ee
Here $\ell$ is the length of the single interval of the Riemann surface $\mc R_{n,1}$. The quasiprimary operator $\Phi_K$ of $CFT^n$ has conformal weight $(h_K,\bar h_K)$, and it is normalized as
\be
\lag \Phi_K(z_1,\bar z_1) \Phi_L(z_2,\bar z_2) \rag_C=\f{\d_{KL}\a_K}{z_{12}^{2h_K}\bar z_{12}^{2\bar h_K}}.
\ee
In the case of two intervals with small cross ratio $x$, as discussed in \cite{Chen:2013kpa,Chen:2013dxa}, the partition function of $CFT^n$ is
\be \label{renyiall}
\tr{\r_A^n} = c_n^2 x^{-\f{c+\bar{c}}{12}(n-\f{1}{n})}\sum_{K}\alpha_K d_K^2 x^{h_K+\bar{h}_K} F(h_K,h_K;2h_K;x) F(\bar h_K,\bar h_K;2\bar h_K;x).
\ee
For a concrete CFT, we need firstly find the quasiprimary operators level by level and determine the corresponding normalization factors $\alpha_K$ and the coefficients $d_K$, then sum over all the contributions according to the relation (\ref{renyiall}). It is easy to see that in the small interval limit, the contributions from the quasi-primiary operators with higher levels are suppressed by power.

In the following we will compute the R\'enyi mutual information in LCFT. Our strategy is to compute it in the normal CFT with extra primary field(s). The essential point is that we must take into account of the quasiprimary operators from the primary field $X$, besides the ones from vacuum Verma module.  At the end of computation we take the $c\to0$ limit to read the mutual information.

Another related subtle issue is on the large central charge limit. In the LCFT at hand, as at least one central charge is zero, the usual large central charge limit seems break down. Nevertheless, in taking $c\to0$ limit, we find that the result could be effectively classified in terms of the power of $1/|b|$, rather than $1/c$. As the anomaly $b$ is finite and inversely proportional to Newton constant $G$, the terms proportional to $1/|b|$ correspond to 2-loop corrections to HRE, while the terms independent of $b$ correspond to 1-loop corrections to HRE. Certainly if the other central charge is non-vanishing, we have to organize the result in the corresponding sector as in usual large central charge limit, and combine the results together.

\subsection{LCFT dual to critical CTMG}

As a warmup, let us first discuss the holographic R\'enyi entropy in the chiral gravity.
The chiral gravity is defined with respect to the Brown-Henneaux boundary conditions (\ref{BHcondition}). As the gauge symmetry at the chiral point is enhanced, we actually have only one massless boundary graviton. Therefore the chiral gravity is conjectured to be dual to  a chiral CFT with only right-moving sector. In this picture, the anti-holomorphic stress-tensor corresponds to the massless graviton. Consequently in computing  the 1-loop R\'{e}nyi entropy using the short interval expansion, we only consider the quasiprimary operators in the anti-holomorphic sector, and find that the R\'enyi mutual information  is just half of the result in \cite{Chen:2013kpa,Chen:2013dxa}. It is in exact agreement with the bulk result, i.e. half of the result in \cite{Barrella:2013wja}.

The CTMG  at critical point with the asymptotic boundary conditions (\ref{logboundary}) includes two massless boundary graviton and one logarithmic mode. It was conjectured to be dual  a logarithmic conformal field theory with the central charge $c=0,\bar{c}\neq0$. Correspondingly on the CFT side, we not only  consider the stress tensors in the left- and right-moving sectors $T(z),\bar{T}(\bar{z})$, but also another pseudo energy momentum tensor $t(z,\bar{z})$.
In order to compute the R\'enyi entropy, we need to know the quasiprimary operators in the theory, including the ones from pseudo energy momentum tensor. However, instead of working with the pseudo energy tensor directly,   we take the picture that the logarithmic conformal field theory is the limit of a normal conformal field theory  as described in the previous subsection. Therefore, we must consider the quasiprimary operators from the primary field $X$, along with the ones from vacuum module.

\begin{table}
  \centering
  \begin{tabular}{|c|c|c|}\hline
    $L_0 $ & quasiprimary operators & degeneracies  \\\hline
    0 & 1 & 1 \\ \hline

    2 & $T_j$ & $n$ \\\hline
    \multirow{2}*{4} & $\mc A_j$ & $n$ \\ \cline{2-3}
       &$T_{j_1}T_{j_2}$ with $j_1 < j_2$ & $\f{n(n-1)}{2}$  \\\hline

    5 & $\mc J_{j_1 j_2}$ with $j_1 < j_2$ & $\f{n(n-1)}{2}$ \\ \hline
      & $\mc B_j$ & $n$  \\ \cline{2-3}
      & $\mc D_j$ & $n$  \\ \cline{2-3}
    6 & $T_{j_1} \mc A_{j_2}$ with $j_1 \neq j_2$ & $n(n-1)$  \\ \cline{2-3}
      & $\mc K_{j_1j_2}$ with $j_1 < j_2$ & $\f{n(n-1)}{2}$      \\ \cline{2-3}
      & $T_{j_1}T_{j_2}T_{j_3}$ with $j_1 < j_2<j_3$ & $\f{n(n-1)(n-2)}{6}$  \\ \hline

    $\cdots$ & $\cdots$ & $\cdots$ \\
    \hline
\end{tabular}
  \caption{Holomorphic quasiprimary operators from vacuum conformal family}\label{vacuum}
\end{table}

The quasiprimary operators from vacuum module has been studied carefully in \cite{Chen:2013kpa,Chen:2013dxa}. Here we just list the holomorphic quasiprimary operators to level six in Table~\ref{vacuum}. In the table, the operators are respectively
\bea
&& \mc A=(TT)-\f{3}{10}\p^2T , \nn\\
&& \mc B=(\p T\p T)-\f{4}{5}(T\p^2T)+\f{23}{210}\p^4T,  \nn\\
&& \mc D=(T(TT))-\f{9}{10}(T\p^2 T)+\f{4}{35}\p^4 T+\f{93}{70c+29} \mc B \nn\\
&& \mc J_{j_1j_2}=T_{j_1}i\p T_{j_2}-i\p T_{j_1}T_{j_2},  \nn\\
&& \mc K_{j_1j_2}=\p T_{j_1} \p T_{j_2}-\f{2}{5} \lt( T_{j_1}\p^2 T_{j_2}+\p^2 T_{j_1}T_{j_2} \rt).
\eea
Their normalization constants $\a_K$'s are respectively
\bea
&& \a_1=1, ~~~
   \a_T=\f{c}{2}, ~~~
   \a_{\mc A}=\f{c(5c+22)}{10},~~~
   \a_{\mc B}=\frac{36c (70 c+29)}{175},  \nn\\
&& \a_{\mc D}=\frac{3 c (2 c-1) (5 c+22) (7 c+68)}{4 (70 c+29)}, ~~~
   \a_{TT}=\f{c^2}{4},   \\
&& \a_{T \mA}=\f{c^2(5c+22)}{20},  ~~~
   \a_{TTT}=\f{c^3}{8}, ~~~
   \a_\mJ=2c^2, ~~~
   \a_\mK=\f{36c^2}{5}, \nn
\eea
and the coefficients $d_K$'s are respectively
\bea
&&  d_1=1, ~~~ d_T=\frac{n^2-1}{12n^2}, ~~~ d_\mA=\frac{(n^2-1)^2}{288 n^4},
~~~ d_\mB=-\frac{(n^2-1)^2 \left(2 n^2(35 c+61)-93\right)}{10368 n^6(70 c+29)},  \nn\\
&& d_\mD=\frac{(n^2-1)^3}{10368 n^6},  ~~~
   d_{TT}^{j_1j_2}=\f{1}{8n^4c}\f{1}{s^4_{j_1j_2}}+\f{(n^2-1)^2}{144n^4},
~~~ d_{T\mA}^{j_1j_2}=\f{n^2-1}{96n^6c}\f{1}{s^4_{j_1j_2}}+\f{(n^2-1)^3}{3456n^6},  \nn\\
&& d_{TTT}^{j_1j_2j_3}=-\f{1}{8n^6c^2}\f{1}{s^2_{j_1j_2}s^2_{j_2j_3}s^2_{j_3j_1}}
                       +\f{n^2-1}{96n^6c} \lt( \f{1}{s^4_{j_1j_2}}+\f{1}{s^4_{j_2j_3}}+\f{1}{s^4_{j_3j_1}} \rt)
                       +\f{(n^2-1)^3}{1728n^6}, \nn\\
&& d_{\mJ}^{j_1j_2}=\f{1}{16n^5 c}\f{c_{j_1j_2}}{s^5_{j_1j_2}},
~~~ d_{\mK}^{j_1j_2}=\f{5}{128n^6c}\f{1}{s^6_{j_1j_2}}-\f{n^2+9}{288n^6c}\f{1}{s^4_{j_1j_2}}-\f{(n^2-1)^2}{5184n^4}.
\eea

\begin{table}
  \centering
  \begin{tabular}{|c|c|c|c|} \hline
  $\#$             & $(L_0, \bar L_0 )$    & quasiprimary operators            & degeneracies     \\ \hline

  4                & $(4+2\e(c),2\e(c))$   & $X_{j_1}X_{j_2}$ with $j_1 < j_2$ & $\f{n(n-1)}{2}$  \\ \hline

  \multirow{2}*{5} & $(5+2\e(c),2\e(c))$   & $\mQ_{j_1 j_2}$ with $j_1 < j_2$  & $\f{n(n-1)}{2}$  \\ \cline{2-4}
                   & $(4+2\e(c),1+2\e(c))$ & $\mR_{j_1 j_2}$ with $j_1 < j_2$  & $\f{n(n-1)}{2}$  \\ \hline

  \multirow{8}*{6} & $(6+2\e(c),2\e(c))$   & $X_{j_1} \mO_{j_2}$ with $j_1 \neq j_2$                                      & ${n(n-1)}$   \\ \cline{2-4}
                   & $(4+2\e(c),2+2\e(c))$ & $X_{j_1} \mP_{j_2}$ with $j_1 \neq j_2$                                      & ${n(n-1)}$  \\ \cline{2-4}
                   & $(5+2\e(c),1+2\e(c))$ & $\mS_{j_1j_2}$ with $j_1<j_2$                                                & $\f{n(n-1)}{2}$  \\ \cline{2-4}
                   & $(6+2\e(c),2\e(c))$   & $\mU_{j_1j_2}$ with $j_1<j_2$                                                & $\f{n(n-1)}{2}$  \\ \cline{2-4}
                   & $(4+2\e(c),2+2\e(c))$   & $\mV_{j_1j_2}$ with $j_1<j_2$                                                & $\f{n(n-1)}{2}$  \\ \cline{2-4}
                   & $(6+2\e(c),2\e(c))$   & $T_{j_1}X_{j_2}X_{j_3}$ with $j_1\neq j_2$, $j_1\neq j_3$ and $j_2<j_3$      & $\f{n(n-1)(n-2)}{2}$ \\ \cline{2-4}
                   & $(4+2\e(c),2+2\e(c))$ & $\bar T_{j_1}X_{j_2}X_{j_3}$ with $j_1\neq j_2$, $j_1\neq j_3$ and $j_2<j_3$ & $\f{n(n-1)(n-2)}{2}$ \\ \cline{2-4}
                   & $(6+3\e(c),3\e(c))$   & $X_{j_1}X_{j_2}X_{j_3}$ with $j_1<j_2<j_3$                                   & $\f{n(n-1)(n-2)}{6}$ \\ \hline

$\cdots$ & $\cdots$ & $\cdots$ & $\cdots$ \\
\hline
\end{tabular}
  \caption{Quasiprimary operators from the conformal family of $X$}
  \label{X}
\end{table}

Apart from the quasiprimary operators of $CFT^n$ constructed by the operators in the  vacuum conformal family, we also need the ones constructed in terms of $X$. To level six the additional quasiprimary operators we need are listed  in Table~\ref{X}.  In the table there is the definition
\be \#=\lim_{c \to 0} \lt( L_0+\bar L_0 \rt).  \ee
For the normal CFT we have
\be
\mO=(TX)-\f{3}{2(2h+1)}\p^2X, ~~~
\mP=(\bar TX)-\f{3}{2(2\bar h+1)}\bar \p^2X,
\ee
with the normalizations
\be
\a_\mO=\f{(2h+1)c+2h(8h-5)}{2(2h+1)}\a_X, ~~~
\a_\mP=\f{(2\bar h+1)\bar c+2\bar h(8\bar h-5)}{2(2\bar h+1)}\a_X.
\ee
For the $CFT^n$ we have
\bea
&& \mQ_{j_1j_2}=X_{j_1}i\p X_{j_2}-i\p X_{j_1}X_{j_2}, ~~~
   \mR_{j_1j_2}=X_{j_1}i\bar\p X_{j_2}-i\bar\p X_{j_1}X_{j_2}, \nn\\
&& \mS_{j_1j_2}=X_{j_1}\p\bar\p X_{j_2}+\p\bar\p X_{j_1} X_{j_2}-\p X_{j_1}\bar\p X_{j_2}-\bar\p X_{j_1}\p X_{j_2},  \nn\\
&& \mU_{j_1j_2}=\p X_{j_1}\p X_{j_2}-\f{h}{2h+1} \lt( X_{j_1}\p^2 X_{j_2} + \p^2 X_{j_1} X_{j_2} \rt), \nn\\
&& \mV_{j_1j_2}=\bar\p X_{j_1}\bar\p X_{j_2}-\f{\bar h}{2\bar h+1} \lt( X_{j_1}\bar\p^2 X_{j_2} + \bar\p^2 X_{j_1} X_{j_2} \rt).
\eea
We have the normalizations
\bea \label{akx}
&& \a_{XX}=i^{4s}\a_X^2, ~~~ \a_\mQ=4h i^{4s}\a_X^2, ~~~ \a_\mR=4\bar h  i^{4s}\a_X^2, ~~~
   \a_{X\mO}=\f{(2h+1)c+2h(8h-5)}{2(2h+1)} i^{4s}\a_X^2, \nn\\
&& \a_{X\mP}=\f{(2\bar h+1)\bar c+2\bar h(8\bar h-5)}{2(2\bar h+1)} i^{4s}\a_X^2, ~~~ \a_\mS=16h\bar h i^{4s}\a_X^2, ~~~
   \a_\mU=\f{4h^2(4h+1)}{2h+1} i^{4s}\a_X^2, \nn\\
&& \a_\mV=\f{4\bar h^2(4\bar h+1)}{2\bar h+1} i^{4s}\a_X^2, ~~~ \a_{TXX}=\f{c}{2} i^{4s}\a_X^2, ~~~ \a_{\bar TXX}=\f{\bar c}{2} i^{4s}\a_X^2, ~~~
   \a_{XXX}= i^{4s}\a_X^3,
\eea
where the factor $i^{4s}=(-1)^{2s}$ aries from the possible sign when $X$ is an fermionic operator. Note that there is always $i^{8s}=1$.
The coefficients $d_K$ for these quasiprimary operators are respectively
\bea \label{dkx}
&& d_{XX}^{j_1j_2}=\f{i^{2s}}{\a_X(2n)^{2\D}}\f{1}{s_{j_1j_2}^{2\D}}, ~~~
   d_\mQ^{j_1j_2}=-d_\mR^{j_1j_2}=\f{i^{2s}}{\a_X(2n)^{2\D+1}}\f{c_{j_1j_2}}{s_{j_1j_2}^{2\D+1}},   \nn\\
&& d_{X\mO}^{j_1j_2}=d_{X\mP}^{j_1j_2}=\f{i^{2s}(n^2-1)}{3\a_X(2n)^{2\D+2}}\f{1}{s_{j_1j_2}^{2\D}},  ~~~
   d_\mS^{j_1j_2}=\f{i^{2s}}{\a_X(2n)^{2\D+2}} \f{c_{j_1j_2}^2}{s_{j_1j_2}^{2\D+2}},   \nn\\
&& d_\mU^{j_1j_2}=\f{i^{2s}}{2h(4h+1)\a_X(2n)^{2\D+2}}\f{(2h+1)(4h+1)-2h(n^2+4h+1)s_{j_1j_2}^2}{s_{j_1j_2}^{2\D+2}},  \nn\\
&& d_\mV^{j_1j_2}=\f{i^{2s}}{2\bar h(4\bar h+1)\a_X(2n)^{2\D+2}}\f{(2\bar h+1)(4\bar h+1)-2\bar h(n^2+4\bar h+1)s_{j_1j_2}^2}{s_{j_1j_2}^{2\D+2}},  \nn\\
&& d_{TXX}^{j_1j_2j_3}=\f{i^{2s}}{\a_X(2n)^{2\D+2}} \lt( -\f{2h}{c} \f{1}{s_{j_1j_2}^2s_{j_1j_3}^2s_{j_2j_3}^{2\D-2}}
                                                         +\f{n^2-1}{3}\f{1}{s_{j_2j_3}^{2\D}} \rt),\nn\\
&& d_{\bar TXX}^{j_1j_2j_3}=\f{i^{2s}}{\a_X(2n)^{2\D+2}} \lt( -\f{2\bar h}{\bar c} \f{1}{s_{j_1j_2}^2s_{j_1j_3}^2s_{j_2j_3}^{2\D-2}}
                                                         +\f{n^2-1}{3}\f{1}{s_{j_2j_3}^{2\D}} \rt),\nn\\
&& d_{XXX}^{j_1j_2j_3}=\f{i^s C_{XXX}}{\a_X^3(2n)^{3\D}}\f{1}{\lt( s_{j_1j_2}s_{j_1j_3}s_{j_2j_3} \rt)^{\D}}.
\eea
Here we have defined $s_{j_1j_2}\equiv\sin\f{\pi(j_1-j_2)}{n}$, $c_{j_1j_2}\equiv\cos\f{\pi(j_1-j_2)}{n}$, $\cdots$ for simplicity. Note that the formulas (\ref{akx}) and (\ref{dkx}) are general and apply to any nonchiral primary operator $X(z,\bar z)$.

Taking the limit $c\to0$, we find
\bea \label{adk2}
&& \a_{XX} \lt( d_{XX}^{j_1j_2} \rt)^2 \to \f{1}{(2n)^8}\f{1}{s_{j_1j_2}^8}, ~~~
   \a_{\mQ} \lt( d_{\mQ}^{j_1j_2} \rt)^2 \to \f{8}{(2n)^{10}}\f{c_{j_1j_2}^2}{s_{j_1j_2}^{10}},  ~~~
   \a_{\mR} \lt( d_{\mR}^{j_1j_2} \rt)^2 \to 0,  \nn\\
&& \a_{X\mO} \lt( d_{X\mO}^{j_1j_2} \rt)^2 \to \f{22(n^2-1)^2}{45(2n)^{12}}\f{1}{s_{j_1j_2}^8},  ~~~
   \a_{X\mP} \lt( d_{X\mP}^{j_1j_2} \rt)^2 \to \f{\bar c (n^2-1)^2}{18(2n)^{12}}\f{1}{s_{j_1j_2}^8},  ~~~
   \a_{\mS} \lt( d_{\mS}^{j_1j_2} \rt)^2 \to 0,  \nn\\
&& \a_{\mU} \lt( d_{\mU}^{j_1j_2} \rt)^2 \to \f{1}{45(2n)^{12}}\f{\lt( 45-4(n^2+9)s_{j_1j_2}^2 \rt)^2}{s_{j_1j_2}^{12}},  ~~~
   \a_{\mV} \lt( d_{\mV}^{j_1j_2} \rt)^2 \to \f{1}{(2n)^{12}}\f{1}{s_{j_1j_2}^{12}},  \nn\\
&& \a_{TXX} \lt( d_{TXX}^{j_1j_2j_3} \rt)^2 x^{2\D+2}  \to \f{x^6}{(2n)^{12}} \lt( \f{1}{\lt( s_{j_1j_2}s_{j_1j_3}s_{j_2j_3} \rt)^4}
                                                           \lt( \f{8}{c}-\f{8}{b} \Big( 1-8\log(2ns_{j_2j_3})+4\log x \Big) \rt) \rt. \nn\\
&& \phantom{\a_{TXX} \lt( d_{TXX}^{j_1j_2j_3} \rt)^2 x^{2\D+2}  \to} \lt.
                                                                     -\f{4(n^2-1)}{3}\f{1}{ s_{j_1j_2}^2s_{j_1j_3}^2s_{j_2j_3}^6 } \rt) \nn\\
&& \a_{\bar TXX} \lt( d_{\bar TXX}^{j_1j_2j_3} \rt)^2 \to \f{\bar c(n^2-1)^2}{18(2n)^{12}} \f{1}{s_{j_2j_3}^8},   \\
&& \a_{XXX} \lt( d_{XXX}^{j_1j_2j_3} \rt)^2 x^{3\D} \to \f{x^6}{(2n)^{12}} \f{1}{\lt( s_{j_1j_2}s_{j_1j_3}s_{j_2j_3} \rt)^4}
                      \lt( -\f{32}{c} \rt.\nn\\
&& \phantom{\a_{XXX} \lt( d_{XXX}^{j_1j_2j_3} \rt)^2 x^{3\D} \to }
                      \lt. +\f{16}{b}\Big( 3-24\log(2n)-8\log(s_{j_1j_2}s_{j_2j_3}s_{j_3j_1})+12\log x \Big)  \rt).  \nn
\eea

Taking into account of all the contributions, we find the R\'enyi mutual information
\be
I_n^{TMG}=I_{n,TMG}^{tree}+I_{n,TMG}^{1-loop}+I_{n,TMG}^{2-loop}+\cdots.
\ee
The tree part is
\bea
&& I_{n,TMG}^{tree}=\frac{\bar c (n-1) (n+1)^2 x^2}{288 n^3}+\frac{\bar c (n-1)(n+1)^2 x^3}{288 n^3}  \nn\\
&& \phantom{I_{n,TMG}^{tree}=}
              +\frac{\bar c (n-1) (n+1)^2 \left(1309 n^4-2 n^2-11\right) x^4}{414720 n^7}  \nn\\
&& \phantom{I_{n,TMG}^{tree}=}
              +\frac{\bar c (n-1) (n+1)^2 \left(589 n^4-2 n^2-11\right) x^5}{207360 n^7}   \\
&& \phantom{I_{n,TMG}^{tree}=}
              +\frac{\bar c (n-1) (n+1)^2 \left(805139 n^8-4244 n^6-23397 n^4-86 n^2+188\right) x^6}{313528320 n^{11}}+O(x^7), \nn
\eea
and this is in accord with the bulk result (\ref{incl}). As $\bar{c}=\frac{3l}{G_N}$, the parts proportional to $\bar{c}$ gives the tree-level contribution in the bulk. It is exactly the same as the one obtained in the pure AdS$_3$ gravity up to order $x^6$. This justifies our argument that in CTMG the classical contribution to the HRE is the same as the one in pure AdS$_3$ gravity.
The 1-loop part is
\bea
&& I_{n,TMG}^{1-loop}=\frac{(n+1) \left(n^2+11\right) \left(3 n^4+10 n^2+227\right) x^4}{2419200 n^7} \nn\\
&& \phantom{I_{n,TMG}^{1-loop}=}
                     +\frac{(n+1) \left(109 n^8+1495 n^6+11307 n^4+81905 n^2-8416\right) x^5}{39916800 n^9}\nn\\
&& \phantom{I_{n}^{1-loop}=}
                     +\frac{(n+1) x^6}{1307674368000 n^{11}} \left(5415533 n^{10}+71680823 n^8+527196884 n^6+3752553304 n^4 \rt. \nn\\
&& \phantom{I_{n}^{1-loop}=} \lt. -625541417 n^2-29929127\right)
                     +O(x^7),
\eea
which agrees exactly with the bulk gravity result (\ref{intmg}).

The 2-loop part is
\bea
&& I_{n,TMG}^{2-loop}=\frac{(n+1) (n^2-4)\big( 1-8\log(2n)+4\log x \big) x^6}{46702656000 n^{11} b}
\left(19 n^8+875 n^6+22317 n^4 \rt. \nn\\
&& \phantom{I_{n,TMG}^{2-loop}=} \lt. +505625 n^2+5691964\right) \nn\\
&& \phantom{I_{n,TMG}^{2-loop}=}
   -\f{x^6}{64n^{12}(n-1)b} \sum_{0\leq j_1 < j_2 <j_3 \leq n-1} \f{\log(s_{j_1j_2}s_{j_2j_3}s_{j_3j_1})}{(s_{j_1j_2}s_{j_2j_3}s_{j_3j_1})^4}  \label{2loopTMG}\\
&& \phantom{I_{n,TMG}^{2-loop}=}
+\frac{(n+1)(n^2-4) \left(19 n^8+875 n^6+22317 n^4+505625 n^2+5691964\right) x^6}{140107968000 n^{11}\bar c}+O(x^7). \nn
\eea
The terms proportional to $1/b$ come from the holomorphic part, and the term proportional to $1/ \bc$ comes from the antiholomorphic part. In other words, in the antiholomorphic sector the expansion is still in powers of $1/\bc$, while in the holomorphic sector the expansion should be in terms of $1/|b|$. Since both $1/\bc$ and $1/|b|$ are proportional to Newton constant $G$, these terms correspond to the 2-loop correction to HRE in the bulk. As usual, when $n=2$, the 2-loop terms are all vanishing, indicating the fact that the dual bulk partition function on genus 1 Riemann surface is exact.  Moreover there appear several novel terms in the holomorphic sector. There is a logarithmic term $x^6\log x$, which is reminiscent of the logarithmic term in the correlators. And there is a term proportional to $\log(2n)$, which is not expected in usual CFT. Besides, the summation involves the logarithmic function as well, and the result cannot be written as a polynomial of $n$. All these terms could be related to the logarithmic nature of CFT.

From our computation, we can figure out that the contribution proportional to the central charge comes only from the vacuum stress tensor. It gives the tree level contribution in the bulk. In LCFT, the operator $X(z,\bar{z})$ is primary, and the corresponding conformal family has no additional contribution at the tree level. However, at the 1-loop level, both vacuum module and primary field contribute, in accord with the fact in the bulk computation both massless graviton and logarithmic mode give 1-loop corrections.

\subsection{LCFT dual to critical CNMG}

\begin{table}
 \centering
  \begin{tabular}{|c|c|c|c|} \hline
  $\#$             & $(L_0, \bar L_0)$    & quasiprimary operators            & degeneracies     \\ \hline

  4                & $(2\e(c),4+2\e(c))$   & $\bX_{j_1}\bX_{j_2}$ with $j_1 < j_2$ & $\f{n(n-1)}{2}$  \\ \hline

  \multirow{2}*{5} & $(1+2\e(c),4+2\e(c))$   & $\bar\mQ_{j_1 j_2}$ with $j_1 < j_2$  & $\f{n(n-1)}{2}$  \\ \cline{2-4}
                   & $(2\e(c),5+2\e(c))$ & $\bar\mR_{j_1 j_2}$ with $j_1 < j_2$  & $\f{n(n-1)}{2}$  \\ \hline

  \multirow{8}*{6} & $(2+2\e(c),4+2\e(c))$ & $\bar X_{j_1} \bar\mO_{j_2}$ with $j_1 \neq j_2$                      & ${n(n-1)}$   \\ \cline{2-4}
                   & $(2\e(c),6+2\e(c))$ & $\bar X_{j_1} \bar\mP_{j_2}$ with $j_1 \neq j_2$                                      & ${n(n-1)}$  \\ \cline{2-4}
                   & $(1+2\e(c),5+2\e(c))$ & $\bar\mS_{j_1j_2}$ with $j_1<j_2$                                                & $\f{n(n-1)}{2}$  \\ \cline{2-4}
                   & $(2+2\e(c),4+2\e(c))$   & $\bar\mU_{j_1j_2}$ with $j_1<j_2$                                                & $\f{n(n-1)}{2}$  \\ \cline{2-4}
                   & $(2\e(c),6+2\e(c))$   & $\bar\mV_{j_1j_2}$ with $j_1<j_2$                                                & $\f{n(n-1)}{2}$  \\ \cline{2-4}
                   & $(2+2\e(c),4+2\e(c))$   & $T_{j_1}\bX_{j_2}\bX_{j_3}$ with $j_1\neq j_2$, $j_1\neq j_3$ and $j_2<j_3$      & $\f{n(n-1)(n-2)}{2}$ \\ \cline{2-4}
                   & $(2\e(c),6+2\e(c))$ & $\bar T_{j_1}\bX_{j_2}\bX_{j_3}$ with $j_1\neq j_2$, $j_1\neq j_3$ and $j_2<j_3$ & $\f{n(n-1)(n-2)}{2}$ \\ \cline{2-4}
                   & $(3\e(c),6+3\e(c))$   & $\bX_{j_1}\bX_{j_2}\bX_{j_3}$ with $j_1<j_2<j_3$                                   & $\f{n(n-1)(n-2)}{6}$ \\ \hline

  $\cdots$ & $\cdots$ & $\cdots$ & $\cdots$ \\
\hline
\end{tabular}
  \caption{Quasiprimary operators from the conformal family of $\bar X$}
  \label{Xb}
\end{table}

For  the CNMG at critical point with the asymptotic boundary conditions (\ref{CNMGlogboundary}), it has two massless gravitons and two logarithmic modes, and holographically dual to a logarithmic conformal field theory with the central charge $c = \bar{c} = 0$. In this case, there are two pseudo energy momentum tensor operators $t(z,\bar{z})$ and $\bar{t}(z,\bar{z})$.
To calculate the partition function of $CFT^n$ in this case, we need to consider not only the quasiprimary operators constructed by the operators of vacuum conformal family and $X$ conformal family, but also the ones constructed by the operators of $\bar X$ conformal family and the ones from the mixing of different conformal families. The quasiprimary operators constructed using the operators in conformal family of $X$ has been listed in Table~\ref{X}, and the ones constructed using the operators in conformal family of $\bX$ is listed in Table~\ref{Xb}. For the normal CFT we have
\be
\bar \mO=(T\bar X)-\f{3}{2(2\bar h+1)}\p^2\bar X, ~~~
\bar \mP=(\bar T\bar X)-\f{3}{2(2 h+1)}\bar \p^2\bar X,
\ee
and for the $CFT^n$ we have
\bea
&& \bar\mQ_{j_1j_2}=\bX_{j_1}i\p\bX_{j_2}-i\p \bX_{j_1}\bX_{j_2}, ~~~
   \bar\mR_{j_1j_2}=\bX_{j_1}i\bar\p \bX_{j_2}-i\bar\p \bX_{j_1}\bX_{j_2}, \nn\\
&& \bar\mS_{j_1j_2}=\bX_{j_1}\p\bar\p \bX_{j_2}+\p\bar\p \bX_{j_1} \bX_{j_2}-\p \bX_{j_1}\bar\p \bX_{j_2}-\bar\p \bX_{j_1}\p \bX_{j_2},  \nn\\
&& \bar\mU_{j_1j_2}=\p \bX_{j_1}\p \bX_{j_2}-\f{\bh}{2\bh+1} \lt( \bX_{j_1}\p^2 \bX_{j_2} + \p^2\bX_{j_1} \bX_{j_2} \rt), \nn\\
&& \bar\mV_{j_1j_2}=\bar\p \bX_{j_1}\bar\p \bX_{j_2}-\f{h}{2h+1} \lt( \bX_{j_1}\bar\p^2 \bX{j_2} + \bar\p^2 \bX_{j_1} \bX_{j_2} \rt).
\eea
Moreover we need to consider the two quasiprimary operators listed in Table~\ref{XXb}.

\begin{table}
 \centering
  \begin{tabular}{|c|c|c|c|} \hline
  $\#$             & $(L_0, \bar L_0)$      & quasiprimary operators                                                        & degeneracies         \\ \hline
  \multirow{2}*{6} & $(4+3\e(c),2+3\e(c))$  & $X_{j_1}X_{j_2}\bX_{j_3}$ with $j_1<j_2$, $j_1 \neq j_3$ and $j_2 \neq j_3$   & $\f{n(n-1)(n-2)}{2}$ \\ \cline{2-4}
                   & $(2+3\e(c),4+3\e(c))$  & $\bX_{j_1}\bX_{j_2}X_{j_3}$ with $j_1<j_2$, $j_1 \neq j_3$ and $j_2 \neq j_3$ & $\f{n(n-1)(n-2)}{2}$ \\ \hline

  $\cdots$ & $\cdots$ & $\cdots$ & $\cdots$ \\
\hline
\end{tabular}
  \caption{The additional quasiprimary operators}
  \label{XXb}
\end{table}

For the quasiprimary operators in Table~\ref{X}, only two relations in (\ref{adk2}) change
\be
\a_{X\mP} \lt( d_{X\mP}^{j_1j_2} \rt)^2 \to 0, ~~~
\a_{\bar TXX} \lt( d_{\bar TXX}^{j_1j_2j_3} \rt)^2 \to 0.
\ee
It can be seen easily that the operators in Table~\ref{Xb} contribute the same as the ones in Table~\ref{X}.

 For the operator $X_{j_1}X_{j_2}\bX_{j_3}$  in Table~\ref{XXb}, we find
\be
\a_{XX\bX}=i^{4s}\a_X^3, ~~~ d_{XX\bX}^{j_1j_2j_3}=\f{i^{3s}C_{XX\bX}}{\a_X^3(2n)^{3\D}}\f{1}{\lt( s_{j_1j_2} s_{j_1j_3} s_{j_2j_3} \rt)^\D},
\ee
and when $c \to 0$ we have
\be
\a_{XX\bX} \lt( d_{XX\bX}^{j_1j_2j_3} \rt)^2 \to 0.
\ee
Similarly when $\bc \to 0$ we have
\be
\a_{\bX\bX X} \lt( d_{\bX\bX X}^{j_1j_2j_3} \rt)^2 \to 0.
\ee
So the operators in Table~\ref{XXb} do not contribute to the R\'enyi entropy.

Taking all the contributions into account, we can read the R\'enyi mutual information
\be
I_n^{NMG}=I_{n,NMG}^{tree}+I_{n,NMG}^{1-1oop}+I_{n,NMG}^{2-loop}+\cdots.
\ee
The tree part is vanishing $I_{n,NMG}^{tree}=0$, as we expected. The 1-loop part is
\bea
&& I_{n,NMG}^{1-loop}=\frac{(n+1) \left(n^2+11\right) \left(3 n^4+10 n^2+227\right) x^4}{1814400 n^7} \nn\\
&& \phantom{I_{n}^{NMG}=}
              +\frac{(n+1) \left(109 n^8+1495 n^6+11307 n^4+81905 n^2-8416\right) x^5}{29937600 n^9} \nn\\
&& \phantom{I_{n}^{NMG}=}
              +\frac{(n+1) x^6}{28740096000 n^{11}}\left(158702 n^{10}+2100642 n^8+15450121 n^6+109973341 n^4 \rt. \nn\\
&& \phantom{I_{n}^{NMG}=}\lt. -18280323 n^2-538483\right)+O(x^7),
\eea
which match the gravity result (\ref{innmg}) exactly. The 2-loop part is
\bea
&& I_{n,NMG}^{2-loop}=\frac{(n+1) (n^2-4)\big( 1-8\log(2n)+4\log x \big) x^6}{23351328000 n^{11} b}
\left(19 n^8+875 n^6+22317 n^4 \rt. \nn\\
&& \phantom{I_{n,TMG}^{2-loop}=} \lt. +505625 n^2+5691964\right) \nn\\
&& \phantom{I_{n,NMG}^{2-loop}=}
   -\f{x^6}{32n^{12}(n-1)b} \sum_{0\leq j_1 < j_2 <j_3 \leq n-1} \f{\log(s_{j_1j_2}s_{j_2j_3}s_{j_3j_1})}{(s_{j_1j_2}s_{j_2j_3}s_{j_3j_1})^4}
   +O(x^7),
\eea
which is the double of the 2-loop result in the holomorphic sector in (\ref{2loopTMG}).

\section{Conclusion and discussion}\label{s4}

In this paper we investigated the  R\'enyi entropy of two disjoint intervals with small cross ratio $x$ in the AdS$_3$/LCFT$_2$ correspondence. The quantum gravity in AdS$_3$ is defined with respect to the asymptotic boundary conditions. For CTMG at the critical point, we may impose the Brown-Henneaux boundary conditions or the logarithmic boundary conditions to include or exclude the logarithmic mode. We showed that the classical actions of the gravitational configurations for the R\'enyi entropy are the same as the ones in pure AdS$_3$ gravity, and computed carefully the 1-loop corrections from various fluctuations in both the chiral gravity and the log gravity. For the CNMG at the critical point, there are different boundary conditions to allow one or two logarithmic modes. As the central charges are vanishing, the classical gravitational action are expected to be vanishing, but the 1-loop corrections could be computed in various cases. In computing the 1-loop R\'enyi entropy, we used the method of Schottky uniformization and summing over the representative of primitive conjugacy classes of Schottky group.

The other part of this work was to compute the 1-loop R\'{e}nyi entropy in LCFT. In the cases at hand, the central charge of the LCFT in at least one sector is vanishing.  Such LCFT could be taken as a limit of ordinary CFT. In doing so, another primary operator $X(z,\bar{z})$ with conformal weight $(2+\eps(c),\eps(c))$ has to be introduced, if the central charge in holomorphic sector is vanishing. Therefore, in  discussing the OPE of the twist operators in the short interval limit, we must take into account of the quasiprimary operators from this primary operator, besides the usual ones from vacuum Verma module. We constructed all the quasiprimary operators up to level 6 and
computed their contributions to the R\'enyi entropy. We found that the contributions proportional to the central charges come only from the vacuum Verma module. In the chiral gravity and log gravity case, these contributions are the same as the ones in pure AdS$_3$ gravity, as we expected. The subleading corrections that are independent of the central charges include the contributions from both the vacuum module and primary operators. To order $x^6$, they are in exact match with the gravitational result in all the cases. These agreements provide further support for the massive gravity/CFT correspondence.

It is remarkable that the small interval expansion in our discussion has reached to order $x^6$. Our motivation is two-fold. On the CFT side, the possible 2-loop corrections appear firstly at order $x^6$. This is of particular interest for the LCFT with $c=0$ as naive 2-loop correction is proportional to $1/c$ and thus might be divergent.
On the gravitational side, the  massless gravitons and logarithmic modes have the same contributions to order $x^5$. Their difference appears  at order $x^6$ as well. Therefore the exact agreement between two sides at 1-loop level at order $x^6$ is highly nontrivial.
Moreover, our investigation also shows new feature in the possible 2-loop correction to HRE in the gravity with logarithmic mode. First of all, even though the central charge $c$ is vanishing, the 2-loop correction is not divergent. In this case, the expansion parameter is not $1/c$, but instead another new parameter $1/|b|$ which is finite and proportional to the Newton constant $G$. Secondly, there are some novel terms appearing in the 2-loop contributions. One such term is proportional to $x^6\log x$, which is reminiscent of the logarithmic term in the correlator in LCFT. It would be nice to understand these terms from direct computation in gravity.

It would be illuminating to reconsider the tree-level contribution to HRE from our study. On the CFT side, such contribution comes purely from the vacuum Verma module, as the information on the central charges is encoded in the stress tensors. Therefore it could be read easily with the computations in \cite{Chen:2013kpa,Chen:2013dxa}. Actually such a treatment applies to all kinds of CFT, including the cases with different left- and right- central charges. In usual CFT, the left- and right- sectors are decoupled and their contributions from the vacuum Verma module are quite similar. As a result, one find that the tree-level contributions always take the similar form, up to the sum of the central charges. This has several nontrivial implications if gravity/CFT correspondence is correct. Firstly, for the AdS$_3$ vacuum in various 3D gravity theories, the classical R\'enyi entropies should differ from the ones in pure AdS$_3$ only by an overall factor, confirming our conclusion in Section \ref{s2}.  Secondly, for the warped AdS$_3$ vacuum, the holographic R\'enyi entropies should be proportional to the ones for pure AdS$_3$. From gravitational point of view, there is no good reason to believe this indication. Therefore this raises a serious challenge to the warped AdS$_3$/CFT$_2$ correspondence.


In  \cite{Gaberdiel:2010xv}, the 1-loop thermal partition function of the CTMG can be expressed as
\be Z_{TMG} = Z_{LCFT}^0 + \mbox{multi-particle contribution}. \ee
Because of the pseudo energy tensor $t(z,\bar{z})$, the logarithmic CFT for CTMG is not chiral and has multi-particle contribution.
The $Z_{LCFT}^0$ is
\be
Z_{LCFT}^0 = \left(\prod_{n = 2}^{\infty}\f{1}{|1-q^n|^2}\right)\left(1+\f{q^2}{|1-q|^2}\right).
\ee
 After we expand this one and compare with (\ref{partitionTMG}), we find that the multi-particle contribution
appears from order $q^4$. In the view point of R\'{e}nyi entropy, beyond the order $x^8$, the multi-particle contribution appears. Therefore this opens another window to study LCFT.

The recent study in \cite{Chen:2013dxa,Perlmutter:2013paa} discussed the HRE for the CFT with $W$ symmetry. In this case, computing HRE should include the higher spin fluctuations. In \cite{Chen:2011vp,Chen:2011yx}, the topologically massive higher spin gravity has been constructed, it would be nice to investigate the holographic R\'{e}nyi entropy in this case.

\vspace*{1cm}
\noindent {\large{\bf Acknowledgments}} \\
BC would like to thank Feng-Li Lin and Bo Ning for valuable discussions. BC also thanks NCTS, Taiwan for hospitality during the course of this work.
FYS would like to thank Sean Hartnoll for valuable correspondence.
We thank Matthew Headrick for his Mathematica code Virasoro.nb that could be downloaded at his personal homepage \url{http://people.brandeis.edu/~headrick/Mathematica/index.html}.
The work was in part supported by NSFC Grant No.~11275010, No.~11335012 and No.~11325522.
JJZ was also in part supported by the Scholarship Award for Excellent Doctoral Student granted by the Ministry of Education of China.
\vspace*{1cm}

\begin{appendix}

\section{Some useful formulas} \label{sa}

In the appendix we give some formulas that are used in our calculation. We define
\be
f_m=\sum_{j=1}^{n-1}\f{1}{ \lt( \sin\f{\pi j}{n} \rt)^{2m}},
\ee
and explicitly we need
\bea
&& f_1=\frac{n^2-1}{3}, ~~~ f_2=\frac{(n^2-1) \left(n^2+11\right)}{45} , ~~~
f_3=\frac{(n^2-1)  \left(2 n^4+23 n^2+191\right)}{945} ,  \nn\\
&& f_4=\frac{(n^2-1) \left(n^2+11\right) \left(3 n^4+10 n^2+227\right)}{14175},  \nn\\
&& f_5=\frac{(n^2-1) \left(2 n^8+35 n^6+321 n^4+2125 n^2+14797\right)}{93555}, \\
&& f_6=\frac{(n^2-1) \left(1382 n^{10}+28682 n^8+307961 n^6+2295661 n^4+13803157 n^2+92427157\right)}{638512875}. \nn
\eea
The above formulas are useful because they often appear in the following summations
\bea
&& \sum_{0\leq j_1 <j_2 \leq n-1} \f{1}{s^{2m}_{j_1j_2}}=\f{n}{2}f_m,  \\
&& \sum_{0\leq j_1 <j_2<j_3 \leq n-1} \lt( \f{1}{s^{2m}_{j_1j_2}} +\f{1}{s^{2m}_{j_2j_3}} +\f{1}{s^{2m}_{j_3j_1}} \rt)=\f{n(n-2)}{2}f_m. \nn
\eea
There are also several other useful summation formulas
\bea
&& \sum_{0\leq j_1 <j_2<j_3 \leq n-1} \f{1}{s^2_{j_1j_2}s^2_{j_2j_3}s^2_{j_3j_1}}=
   \frac{n\lt(n^2-1\rt)\lt(n^2-4\rt) \left(n^2+47\right)}{2835},  \nn\\
&& \sum_{0\leq j_1 <j_2<j_3 \leq n-1} \f{1}{s^4_{j_1j_2}s^4_{j_2j_3}s^4_{j_3j_1}}=
   \frac{n\lt(n^2-1\rt)\lt(n^2-4\rt)}{273648375}\left(19 n^8+875 n^6+22317 n^4 \rt. \nn\\
&& \phantom{\sum_{0\leq j_1 <j_2<j_3 \leq n-1} \f{1}{s^4_{j_1j_2}s^4_{j_2j_3}s^4_{j_3j_1}}=} \lt.+505625 n^2+5691964\right) ,  \\
&& \sum_{0\leq j_1 <j_2<j_3 \leq n-1} \f{1}{s^2_{j_1j_2}s^2_{j_2j_3}s^2_{j_3j_1}}
   \lt( \f{1}{s^4_{j_1j_2}}+\f{1}{s^4_{j_2j_3}}+\f{1}{s^4_{j_3j_1}} \rt)=
   \frac{n (n^2-1) (n^2-4)}{467775}\left(3 n^6+125 n^4 \rt.  \nn\\
&& \phantom{\sum_{0\leq j_1 <j_2<j_3 \leq n-1} \f{1}{s^2_{j_1j_2}s^2_{j_2j_3}s^2_{j_3j_1}}\lt( \f{1}{s^4_{j_1j_2}}+\f{1}{s^4_{j_2j_3}}+\f{1}{s^4_{j_3j_1}} \rt)=}
   \lt. +1757 n^2+21155\right),  \nn\\
&& \sum_{0\leq j_1 <j_2<j_3 \leq n-1} \lt( \f{1}{s^4_{j_1j_2}s^4_{j_2j_3}}+\f{1}{s^4_{j_2j_3}s^4_{j_3j_1}}+\f{1}{s^4_{j_3j_1}s^4_{j_1j_2}} \rt)=
\frac{2n (n^2-1) (n^2-4) \left(n^2+11\right) \left(n^2+19\right)}{14175}. \nn
\eea

\end{appendix}



\providecommand{\href}[2]{#2}\begingroup\raggedright\endgroup

\end{document}